\pdfoutput=1
\documentclass[notitlepage]{revtex4-1}
\usepackage[T1]{fontenc}
\usepackage{units}
\usepackage{textcomp}
\usepackage{bm}
\usepackage{amsmath}
\usepackage{amssymb}
\usepackage{graphicx}
\usepackage{esint}
\usepackage{float}
\usepackage{bbold}

\providecommand{\tabularnewline}{\\}

\begin{document}

\title[Signatures of delocalization]
  {Signatures of Exciton Delocalization and Exciton-Exciton Annihilation in Fluorescence-Detected Two-Dimensional Coherent Spectroscopy}

\author{Pavel Mal\'{y}$^{1,2}$ and Tom\'{a}\v{s} Man\v{c}al$^{1}$}\email{mancal@karlov.mff.cuni.cz, maly@karlov.mff.cuni.cz}

\affiliation{$^{1}$Faculty of Mathematics and Physics, Charles University, Ke Karlovu
5, 12116 Prague, Czech Republic, $^{2}$Faculty of Science, Vrije Universiteit Amsterdam, De Boelelaan
1081, 1081HV Amsterdam, The Netherlands}

\begin{abstract}
We present calculations of the fluorescence-detected coherent two-dimensional (F-2DES) spectra of a molecular heterodimer. We compare how the F-2DES technique differs from the standard coherently detected two-dimensional (2DES) spectroscopy in measuring exciton delocalization. We analyze which processes contribute to cross-peaks in the zero-waiting-time spectra obtained by the two methods. Based strictly on time-dependent perturbation theory, we study how in both methods varying degree of cancellation between perturbative contributions gives rise to cross-peaks, and identify exciton annihilation and exciton relaxation contributions to the cross-peak in the zero-waiting-time F-2DES. We propose that time-gated fluorescence detection can be used to isolate the annihilation contribution to F-2DES both to retrieve information equivalent to 2DES spectroscopy and to study annihilation contribution itself.
\end{abstract}

\maketitle

Over the past decade, many variants of multi-dimensional spectroscopic techniques have been developed and new methods for data analysis have been added. One of the recent additions to the arsenal of feasible methods are those which combine phase cycling approach to the relevant data retrieval with incoherent signal detection, such as the fluorescence (FL) detection\cite{Tekavec2007,Karki2014,Draeger2017}. The new detection scheme brings promises of better sensitivity\cite{Goetz2018,Tiwari2018,Karki2018} to certain crucial parameters of molecular aggregates, such as exciton delocalization, and opens questions of contributions from processes related to the detection scheme itself. These promises and questions have to be thoroughly addressed before the method becomes mainstream\cite{Tiwari2018a,Karki2018}. In this paper, we put the theory of fluorescence-detected two-dimensional electronic spectroscopy (F-2DES) into the context of the well-known theory of the standard coherently detected two-dimensional electronic spectroscopy (2DES).
We view the processes of the signal generation in F-2DES as a new degree of freedom which represents both an opportunity for development of new sensitive experiments and a potential source of difficulties in comparing the new method to established spectroscopic techniques. 

Two theoretical frameworks stand out as indispensable foundations for various models aimed at explaining recent experiments, namely {\it Frenkel exciton model} \cite{VanAmerongen2000, May2000} and the {\it perturbative response function theory} of non-linear spectroscopy \cite{Mukamel1995Book,Valkunas2013}. While they are, obviously, approximate theories, some of their essential approximations represent core features of the photo-induced physics of molecular aggregates. For instance, the lack of orbital overlap between neighboring molecules, required in Frenkel exciton model, is one of the features of natural photosynthetic antennae, which prevents excitation quenching related to formation of charge transfer states\cite{Beddard1976}. 
Similarly, response function theory of the third order explains well the typical time-resolved experiments 
including coherent two-dimensional electronic spectroscopy. In these experiments, it is crucial to ensure that the signal depends on the third power of the external field to prevent unwanted higher order signal. 
Thus the order of perturbation is not a limitation of theory, but rather a constitutive feature of the experiment. 

\begin{figure}
\begin{center}
 		\includegraphics[width=3.42in]{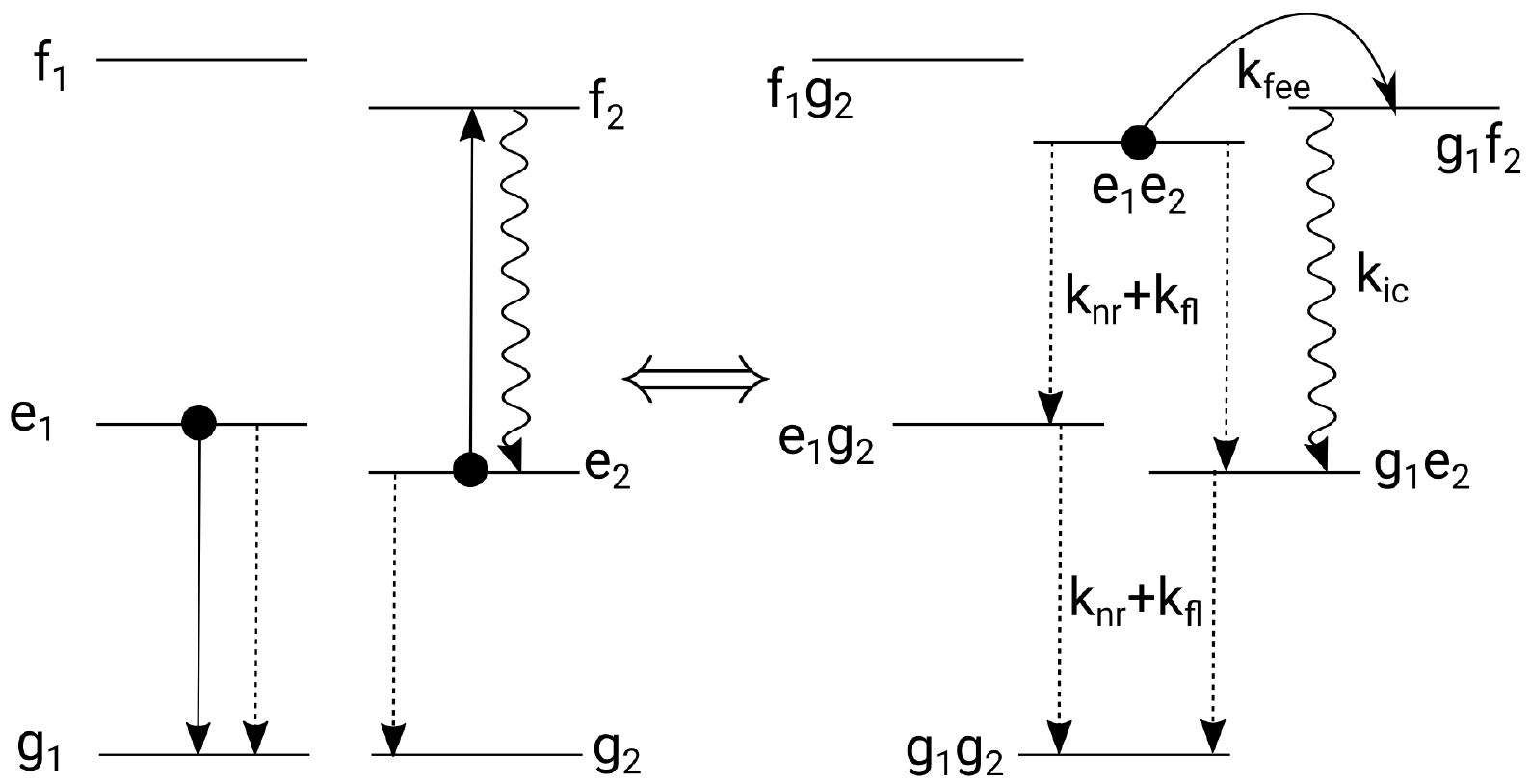}
		\caption{\label{fig:model} Three band model of an effective molecular dimer with exciton-exciton annihilation. Full lines: double-exciton-to-higher-exciton transfer, wiggly line: internal conversion, dashed lines: radiative and non-radiative relaxation.}
		\end{center}
\end{figure}

Exciton delocalization is one of the crucial features of efficient energy transfer in photosynthetic antennae\cite{Strumpfer2012}. Correspondingly, one of the most valuable features of the 2DES spectrum is that the presence of a cross-peak between electronic states of a molecular aggregate at the waiting time $t_2 = 0$ is a direct witness of exciton delocalization. Similar relation between cross-peaks and delocalization has been expected in F-2DES method\cite{Karki2018}. 
Processes occurring on longer time scale than the one defined by the four initial pulses of the experiment, play a significant role in establishing the measured signal in F-2DES. In order to study how is the exciton delocalization reflected in F-2DES, we include into our considerations processes of exciton-exciton (e-e) annihilation and single exciton energy relaxation that occur before fluorescence emission. To this end, in a slight extension of the usual Frenkel exciton model, we consider a dimer of three-level molecules. 
The state diagram is presented in Fig. {\ref{fig:model}}. The eigenstates form three bands: the ground-state $|g\rangle$, the single exciton band $|e\rangle$ (formed from linear combinations of states such as $|n\rangle_1 = |e_{n}\rangle\prod_{k\neq n}^{N}|g_{k}\rangle$), and the double exciton band. The latter contains not only the true double-exciton states $|ee\rangle$, i.e. linear combinations of states $|nm\rangle = |e_n\rangle|e_m\rangle\prod_{k\neq n,m}|g_k\rangle$, but also the manifold $|f\rangle$ of higher excited states $|n\rangle_2 = |f_n\rangle\prod_{k\neq n}|g_k\rangle$ of the pigments. Above we denoted $|e_n\rangle$ the first excited state and $|f_n\rangle$ the second excited state of the n$^{th}$ molecule. In our model we assume optically allowed transitions $|g_n\rangle \rightarrow |e_n\rangle$ and $|e_n\rangle \rightarrow |f_n\rangle$. Transition $|g_n\rangle \rightarrow |f_n\rangle$ is assumed forbidden (or off-resonant with the pulses) for simplicity. Without much loss of generality, we assume that the energy gaps $\hbar\omega^{(n)}_{fe}=\epsilon^{(n)}_f-\epsilon^{(n)}_e$ between the states $|f_n\rangle$ and $|e_n\rangle$ are suitably larger than the gaps $\hbar\omega^{(n)}_{eg} = \epsilon^{(n)}_e-\epsilon^{(n)}_g$ between states $|e_n\rangle$ and $|g_n\rangle$ so that resonance interaction between transitions of the types $|g\rangle \rightarrow |e\rangle$ and $|e\rangle \rightarrow |f\rangle$ residing on neighbouring molecules does not lead to any significant delocalization. 

It is important to note, that it is the energy transfer between these two types of transitions which forms the first step of exciton-exciton annihilation process. We describe this process by energy transfer rates. In a multichromophoric system, once two excitons are created, they migrate and, eventually, meet at coupled pigments, where they fuse into one of the $f$ states in a process $|ee\rangle \rightarrow |f\rangle$. In our model, the effective rate $k_{fee}$ of this process includes both the exciton migration and the first annihilation step. The transfer $|ee\rangle \rightarrow |f\rangle$ occurs uphill, with thermodynamics dictating a faster backward rate. However, it is known that due to internal conversion processes, the $|f_n\rangle$ states rapidly convert into $|e_n\rangle$ states (often on sub-100 fs time scale). We require the backward rate to be slower than the internal conversion process, so that the prevailing process is the one of slow uphill transfer and efficient internal conversion. This is our model of exciton-exciton annihilation. 

F-2DES experiment has been described in detail in literature\cite{Tekavec2007,Draeger2017,Tiwari2018a}. Four short laser pulses with well defined phases $\phi_1$, $\phi_2$, $\phi_3$ and $\phi_4$, respectively, and well defined delays $t_1$, $t_2$ and $t_3$ between first and second, second and third, and third and fourth pulses, respectively, are applied to a sample containing the studied molecular system, and the fluorescence as a function of the time delays and phases is recorded. The signal dependence on the pulse phases is numerically processed in such a way that a signal component equivalent in its phase dependence (on the external pulses) to the signal obtained in the standard 2DES measurement \cite{Brixner2004} is obtained. While in standard measurement, this signal is distinguished from other signals by its spatial direction 
\cite{Brixner2004}, in the F-2DES, equivalent signal has to be disentangled from the rest of the fluorescence detected response by its phase signature. This is done either directly by filtered lock-in detection\cite{Tekavec2007} or numerically\cite{Draeger2017}. The same phase cycling method can be used in 2DES measurements\cite{Tan2008}, for comparison of various 2DES implementations see Ref.\cite{Fuller2015}. 

\begin{figure}
\begin{center}
 		\includegraphics[width=3.42in]{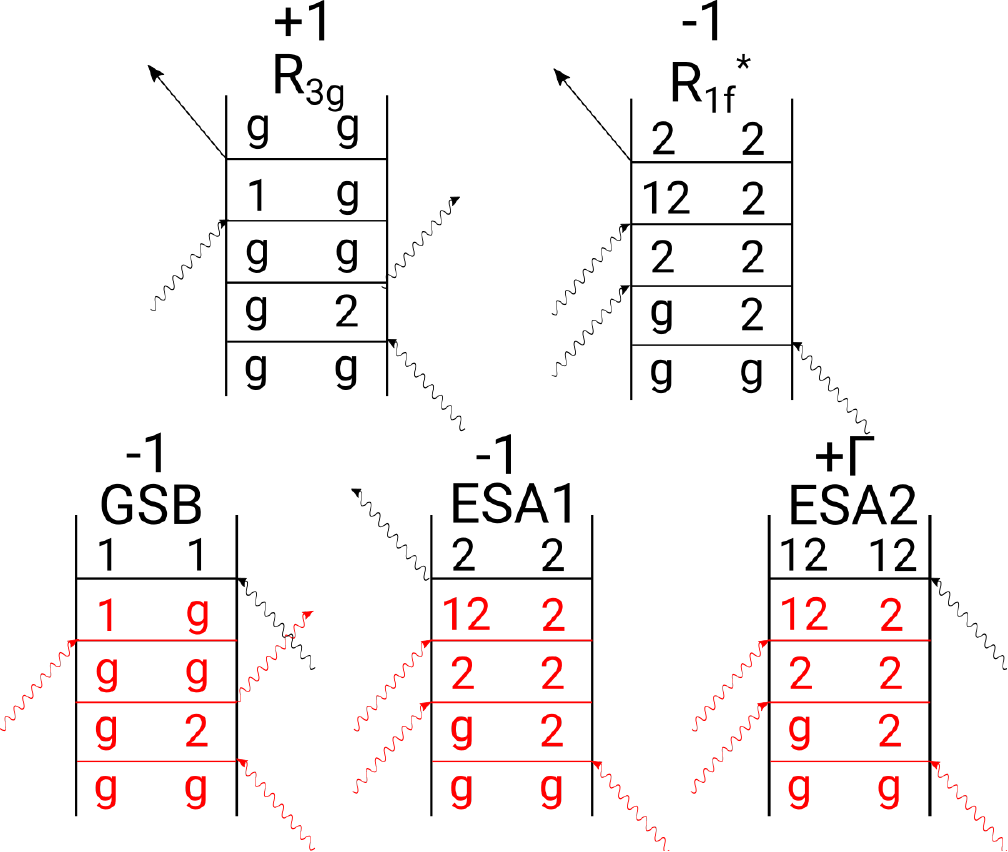}
		\caption{\label{fig:pathways} Part A) Non-linear response functions corresponding to a cross-peak signal in the 2DES experiment (diagrams $R_{3g}$ and $R_{1f}^{*}$). Part B) Non-linear response functions corresponding to the cross-peak signal in F-2DES experiment (diagrams ESA1, ESA2 and GSB). All diagrams are denoted by the sign of their contribution, which is obtained as $-1^{u}$, where $u$ is the number of arrows on the left-hand-side of the diagram. In F-2DES diagrams we distinguish their portion which is identical to the 2DES diagrams by red color, $GSB\leftrightarrow R_{3g}$, $ESA1\leftrightarrow R_{1f}^{*}$. Analogical set of pathways can be written which has an coherence $1\;2$ in the population time, the GSB pathway changes into a stimulated emission one. }
\end{center}
\end{figure}

Assuming quasi-resonant excitation and the limit of ultra-short laser pulses, the non-linear response of the system is proportional (up to a phase factor) to a sum of non-linear response functions of the third order \cite{Mukamel1995Book}. In Fig. \ref{fig:pathways}, we present double-sided Feynman diagrams responsible for a cross-peak on a specific spectral position in both the 2DES and F-2DES spectra. We notice that the diagrams for the 2DES are conventionally closed by the transition dipole operator acting from the left (denoted by the straight line). 
In the case of F-2DES, the diagrams are closed by interaction with the fourth laser pulse, either on the left- or right-hand side of the Feynman diagram. The position of the arrow determines the sign of the contribution (reflects one part of a commutator). We can see that both ESA1 and ESA2 diagrams of the F-2DES experiment are descendants of the diagram $R_{1f}^{*}$ of the 2DES experiment. The theoretical expressions represented by these diagrams are presented in the SI.

\begin{table}
\begin{centering}
\begin{tabular}{|c|c|}
\hline 
quantity & value\tabularnewline
\hline 
\hline 
$\hbar\omega_{eg}^{(1)}$, $\hbar\omega_{eg}^{(2)}$ & $12000$ cm$^{-1}$, $12500$ cm$^{-1}$\tabularnewline
\hline 
$\hbar\omega_{fe}^{(1)}=\hbar\omega_{fe}^{(2)}$ & $13500$ cm$^{-1}$\tabularnewline
\hline 
$k_{fee}\;\left(\approx k_{A}\right)$ & $\frac{1}{10\text{ ps}}$\tabularnewline
\hline 
$k_{ic}$ & $\frac{1}{100\text{ fs}}$\tabularnewline
\hline 
$k_{nr}$, $k_{fl}$ & $\frac{1}{4.3\text{ ns}}$, $\frac{1}{10\text{ ns}}$\tabularnewline
\hline 
$\phi_{fl}=\frac{k_{fl}}{k_{fl}+k_{nr}}$ & $0.3$\tabularnewline
\hline 
$\mu_{eg}^{(1)}=\mu_{fe}^{(1)}$ & $\frac{4.0D}{\sqrt{10}}\left(3,1,0\right)$\tabularnewline
\hline 
$\mu_{eg}^{(2)}=\mu_{fe}^{(2)}$ & $\frac{3.3D}{\sqrt{10}}\left(3,-1,0\right)$\tabularnewline
\hline 
$d_{12}$ & $\in\left[6,50\right]\AA$\tabularnewline
\hline 
$T$ & 300K\tabularnewline
\hline 
$\lambda$, $\tau_{c}$ & $50$ cm$^{-1}$, 100 fs\tabularnewline
\hline 
\end{tabular}
\par\end{centering}
\caption{System parameters for calculations\label{tab:SysParams}}

\end{table}

The rules for conversion of the diagrams of the 2DES measurement into the ones involved in F-2DES are the following: Each conventional Liouville pathway which ends in the ground state in 2DES experiment, yields a single pathway ending in the excited state in F-2DES. The sign of the diagram is inverted, i.e. it is minus in F-2DES\footnote{Concerning the diagram sign, there is a pre-factor of $i$ for each interaction with the pulse. The F-2DES thus has $i^{4}$, while 2DES $i^{3}$. However, the 2DES signal field is connected to the 3rd order polarization as $E_{sig} \sim iP^{3}$, so the $i^{4}$ pre-factor is the same for both detection techniques}. 
Each Liouville pathway which ends in the excited state in 2DES (and which reaches the two-exciton manifold) yields two pathways in F-2DES: one with an inverted sign with respect to the corresponding 2DES pathway, i.e. yielding a positive contribution, and one keeping its sign, i.e. yielding a negative contribution. The former of the F-2DES pathways ends in double exciton manifold, and the latter ends in the single exciton manifold. 

Unlike in 2DES, where the signal is produced in a stimulated processes, in F-2DES measurement, the signal is generated by spontaneous processes which start from the excited state prepared by the four pulses of the F-2DES sequence. The processes following the four-fold interaction with the pulses involve quasi-equilibration within and between the excited state manifolds, including radiative and non-radiative energy relaxation and e-e annihilation. This system dynamics determines the contribution of the individual Liouville-space pathways to the measured F-2DES spectrum. 
Due to the timescale and incoherent nature of the fluorescence emission, it suffices to consider the evolution of the system eigenstates. As the intraband energy relaxation is typically orders of magnitude faster than the interband processes, we will consider the $|e\rangle$  and $|ee\rangle$ manifolds being in quasi-stationary thermal equilibrium.  As we show in the SI, this leads to a set of linear kinetic rate equations for the band populations.  

\begin{figure}[ht]
\begin{center}
 		\includegraphics[width=7in]{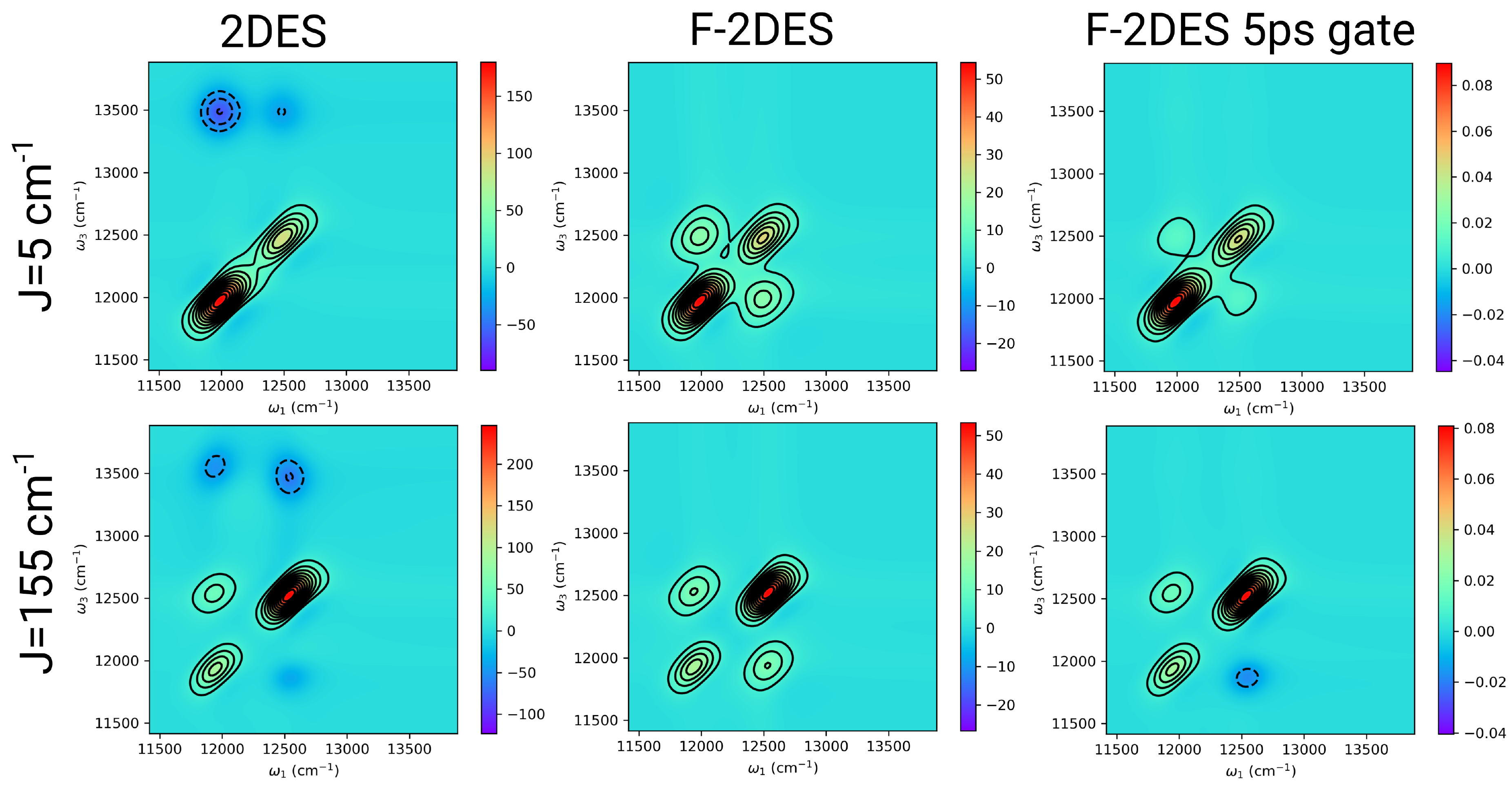}
		\caption{\label{fig:spectra} Simulated 2DES and F-2DES spectra for system with weak (top) and strong (bottom) resonance coupling. Depicted are 2DES (left), F-2DES (middle), and fast-gated F-2DES (right). The spectra were calculated and directly plotted by the QuantaRhei package\cite{quantarhei2018}.}
		\end{center}
\end{figure}

\begin{figure}[H]
\begin{center}
 		\includegraphics[width=7in]{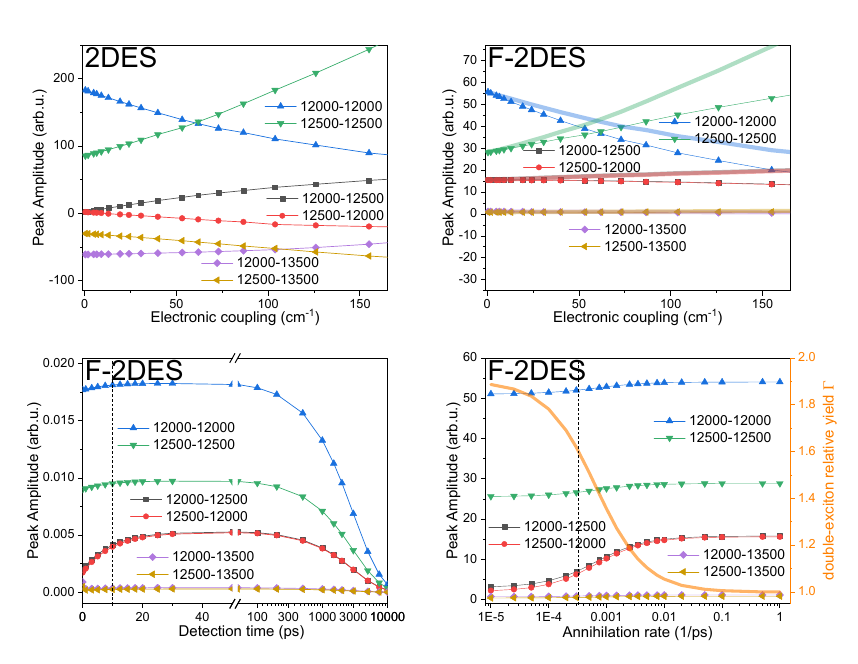}
		\caption{\label{fig:PeakAmplitudes} Dependence of the peak amplitude on the resonance coupling $J$ (top), FL detection time (bottom left, note the logarithmic axis after 50 ps, the dashed line indicates the annihilation time of $10$ ps) and annihilation rate (bottom right, note the logarithmic horizontal axis, the dashed line indicates the excitation lifetime of 3 ns). The labels denote the peak assignment ($w_{1}\;w_{3}$), top pair: diagonal peaks, middle pair: cross peaks, bottom pair: ESA cross peaks. For direct comparison between the 2DES and F-2DES, the latter spectra amplitude can be 'corrected' for one-exciton oscillator strength redistribution (shaded lines top right). In the annihilation rate dependence, bottom right, we give also the relative yield $\Gamma$ of the ESA2 pathway ending in the double-excited state.}
		\end{center}
\end{figure}

From the pathways in Fig. \ref{fig:pathways}, we can infer some features of the cross-peaks in 2DES and F-2DES. In traditional 2DES, there are two pathways contributing to the 2-1 cross-peak, largely cancelling each other out. When the molecules are weakly-coupled (or independent), the cancellation is perfect. However, when the coupling is stronger, the transition-dipole strength redistributes between the states, as seen at the diagonal peaks in Figs. (\ref{fig:spectra}) and (\ref{fig:PeakAmplitudes}). As a result, the cancellation is not complete and it can be shown for molecules with the same oscillator strength, that, in the leading term, the cross peak amplitude is proportional to $\frac{2J}{\Delta E}$. The cross-peak amplitude thus provides good information about the exciton delocalization \footnote{We remind the reader that the mixing angle for a dimer is given by $tan2\theta=\frac{2J}{\Delta E}$.}. 

In the F-2DES there are three Liouville pathways contributing to the cross-peak. The GSB and ESA1 have the same (negative) signs, while ESA2 bears the opposite sign. 
The contribution of the GSB and ESA1 pathways is given by the fluorescence rate of the equilibrated one-exciton manifold, $k_{fl}^{e}$; let us take these pathways to contribute by one photon to the signal. The ESA2 contribution depends on the fate of two-exciton state, signified by the emission rate $k_{fl}^{ee}$ and the annihilation rate, $k_{A}\approx k_{fee}$. In the case that the molecules are independent, the ESA2 pathway contributes by two photons and the cross-peak vanishes, similarly to 2DES. It is important to note that the cross-peak vanishes, in the response theory, only when all Liouville pathways are taken into account. One has to resist the temptation to remove certain pathways, e.g. GSB pathways involving two independent molecules, based on seemingly physical arguments \cite{Karki2018}. Individual Liouville pathways do not correspond to real physical processes, but rather to components of a perturbation series.

In the case of weakly-coupled molecules (energy transfer, but no delocalization), the annihilation plays the key role. When the signal is emitted/detected before the annihilation occurs, the relative yield of the ESA2 pathway is given by $\frac{k_{fl}^{ee}}{k_{fl}^{e}}$. This ratio differs from 2 in the case of energy equilibration between pigments of unequal oscillator strength (OS). For molecules with similar OS, the double exciton state contributes two photons, and the pathways cancel again, suppressing the cross peak.  When, however, the annihilation is present, it reduces the contribution of the ESA2 pathway, preventing the cancellation and giving rise to the cross peaks. Let us now consider molecules of the same OS and focus on the integrated fluorescence, so that it is sufficient to take the overall yield of the ESA2 pathway $\Gamma=1+\left(1+\frac{k_A}{2\left(k_{nr}+k_{fl}\right)}\right)^{-1}$, $\Gamma\in\left[2,1\right]$. 
Denoting the amount of annihilation $A=2-\Gamma=\frac{k_A}{k_A+2\left(k_{nr}+k_{fl}\right)},\;A\in\left[0,1\right]$, the detection-time-integrated F-2DES spectrum can be expressed
\footnote{This applies only when not considering the ESA2 ending in the 
$|f_n>$ 
states. 
For this ESA2, 
$\Gamma \approx 1$ 
due to the fast internal conversion.
} 
as $S_{F-2DES} \propto -GSB-SE+\left(\Gamma -1\right) ESA$ \cite{Lott2011, Karki2014}. 
For no annihilation, $A=0$, the spectrum corresponds to the traditional 2DES, $S_{2DES}=GSB+SE-ESA$. Furthermore, it can be shown that the cross-peak amplitude is directly proportional to the annihilation $A$ and the F-2DES cross-peaks are a good measure for e-e annihilation. We treat the more general case of finite coupling, partial annihilation and different OS further in the text. The derivation of the dependences presented above can be found in the SI.

To demonstrate quantitatively the properties of F-2DES and its sensitivity to exciton delocalization and e-e annihilation, we compare the F-2DES spectra and their parameter dependence with 2DES. The parameters used for the simulations can be found in Table \ref{tab:SysParams} and the computational details are presented in the SI. In Fig. \ref{fig:spectra} we present a time-zero 2D spectrum of our effective dimer with weakly and strongly coupled molecules. We present the F-2DES spectrum with the same sign convention as in the traditional 2DES.
In traditional, coherently-detected 2DES, two diagonal peaks corresponding to the SE and GSB of the two transitions can be found. Additionally, there is a negative ESA to the $f$ states. In the weak coupling limit, there are no cross peaks and no exciton delocalization. With stronger coupling, cross-peaks appear and the excitons become delocalized. Also the oscillator strength becomes redistributed between the transitions. In the F-2DES there are cross peaks both in the weak and strong coupling limit. Interestingly, there is no ESA to the higher excited $f$ states. This is a result of near-perfect cancellation of the ESA1 and ESA2 pathways due to the rapid internal conversion, as can be seen from Fig. \ref{fig:pathways} and footnote 20.
In the last column of Fig. \ref{fig:spectra} we give F-2DES spectra obtained by gating the fluorescence and detecting only the first 5 ps. Clearly the spectra resemble the traditional 2DES, without the $f$ state ESA. 

To learn more, we study the dependence of the peak amplitude on the system parameters (see Fig. \ref{fig:PeakAmplitudes}). In all panels, the top pair of lines depicts amplitudes of the diagonal peaks, the middle pair amplitudes of the cross-peaks, and the bottom pair amplitudes of the $f$ state ESA peaks. In the top row, the peak amplitude with increasing coupling is plotted for 2DES and F-2DES. The dependence confirms what we asserted before. The diagonal peaks in both 2DES and F-2DES behave in the same way (when corrected on the $k_{fl}^{e}$), the oscillator strength of the transitions is redistributed. For 2DES the cross-peak amplitude increases with the coupling, linearly at first, giving a good measure of the coupling (and the resulting exciton delocalization). In the F-2DES, there is no $f$ state ESA peak. The cross-peak amplitude does not vanish even for negligible coupling and displays only slight increase with the coupling. It is therefore not a measure of delocalization. In order to demonstrate the origin of the cross-peaks and the potential extension of the technique, we investigate the possibility of time-gating the detected fluorescence. In the bottom-left panel of Fig. \ref{fig:PeakAmplitudes}, the evolution of the peak amplitude with the detection time unveils. In the first 50 ps (linear scale) the cross-peak amplitude rises exponentially. Crucially, this happens on the timescale of the e-e annihilation (10 ps, the dashed line). In later times (log scale) the whole spectrum decays with the lifetime of the fluorescence. Finally, in the bottom-right panel, we investigate the dependence on the rate of e-e annihilation. For annihilation much slower than the excitation lifetime (3 ns, the dashed line), there are, again, only weak cross peaks, with amplitude given by the difference of the pigment oscillator strength. When, however, the annihilation becomes comparable to or faster than the excitation relaxation rate, pronounced cross-peaks appear, originating from the presence of annihilation. The cross-peak amplitude clearly anticorrelates with the yield of the double-exciton state, reflecting the cancellation of pathways described above.

In summary, we have presented calculations of F-2DES spectra based on rigorous response function formalism. We demonstrate that while cross-peaks at zero waiting time in traditional 2DES are signatures of exciton delocalization, the same cross-peaks in F-2DES can also stem from exciton-exciton annihilation and exciton relaxation processes.
We propose that time-gating of the measured fluorescence signal provides a direct way to observe interactions in two-exciton band, notably the exciton-exciton annihilation.

\section*{Acknowledgments}
This work was supported by the Czech Science Foundation (GA\v{C}R) grant no. 17-22160S.

\section*{Supporting Information}




\subsection*{Response function formulation}

Four wave mixing spectroscopy can be well described in a perturbative language, as a response of the system to several interactions with the electric field, in the dipole approximation $H_{int}=-\mu E(t)$. In the equations below, for the sake of illustration we consider ultrashort pulses. The system propagation is described by a propagator $\mathcal{U}(t)$.
In this contribution we will focus on 2DES; the formalism and considerations are, however, rather general and can be easily applied to other types of nonlinear spectroscopy. 
In the traditional 2DES, the system interacts with three pulses and the signal arises from the macroscopic polarization: 

\begin{equation}
S_{2DES}(t_{3},t_{2},t_{1}) \propto P^{(3)}(t_{3},t_{2},t_{1}) \propto Tr\{\mu\mathcal{U}(t_{3})[\mu,\mathcal{U}(t_{2})[\mu,\mathcal{U}(t_{1})[\mu,\rho^{(eq)}]]]\}.
\end{equation}

\noindent{}In contrast, in the incoherently detected (also called action) 2DES, such as F-2DES, the system interacts with four pulses. Subsequently, an incoherent signal proportional to the state of the system (typically population of its excited state) is detected: 

\begin{eqnarray}
S_{F-2DES}(t;t_{3},t_{2},t_{1}) \propto Tr_{sys}\{\mathcal{K}\rho^{(4)}(t; t_{3},t_{2},t_{1})\} \nonumber \\
\propto Tr\{\mathcal{K} \: \mathcal{U}(t)[\mu,\mathcal{U}(t_{3})[\mu,\mathcal{U}(t_{2})[\mu,\mathcal{U}(t_{1})[\mu,\rho^{(eq)}]]]]\}.
\end{eqnarray}

\noindent{}By $\mathcal{K}$ we have denoted a projector on the system populations, including weighing by the different contribution of the states to the signal emission (given by e.g. a different emission rate due to different transition oscillator strength). The detection time is typically not resolved and the time $t$ is integrated over. In both the coherent and incoherent detection cases, the signals can be expressed by four-point correlation functions of the system dipole, and they are therefore composed from the same components (also called Liouville pathways, following the evolution of the density matrix in Liouville space). However, two crucial differences arise from the different experimental realization. First, the additional interaction with the fourth pulse in the incoherently detected 2DES is represented by two terms from a commutator, while the calculation of the third order polarization involves only one term. As a result, the individual pathways contribute with different signs to the overall signal. Moreover, there is one additional two-exciton pathway in the incoherently-detected 2DES. Second, while in the coherently-detected 2DES the signal polarization field is created as an immediate response to the action of the three pulses, in the incoherently-detected 2DES this is not the case. After the action of the four pulses, the system evolves freely while emitting the observed incoherent signal. As this evolution can proceed differently for the different pathways, it can substantially modify the measured 2D spectrum. This time dependence of the nonlinear signal also opens possibilities for an additional time-resolved detection.  

\subsection*{System dynamics after the interaction with the pulses \label{sec:System-dynamics}}

In order to understand F-2DES spectrum, we need to model the processes occurring after the fourth pulse. We assume that the dynamics in the two-exciton band can be described by considering diagonal elements of the reduced density matrix in the corresponding approximate eigenstate basis, and that relaxation rate description is appropriate.  
We postulate three types of rates, namely internal conversion rates $k^{(n)}_{ef}$ for each molecule, exciton-annihilation rates $K^{(\rm{ann})}_{f_{n},nm}$ which describes energy transfer from a state $|nm\rangle = |e_n\rangle |e_m\rangle \prod_{k\neq n,m}|g_k\rangle$ in which molecules $n$ and $m$ are excited into excited state $|f_{n}\rangle$ on the molecule $n$, and single exciton energy transfer rates $\gamma_{\alpha\beta}$. We assume delocalization in the single-exciton band, i.e. states $|n\rangle_1$ are not (even approximate) eigenstates of the Hamiltonian. The actual eigenstates are delocalized according to $|\alpha\rangle = \sum_{n}c_{\alpha n}|n\rangle_1$. Because we assume no delocalization between the $f$-type- and the double-excited states, and because we are in need of an effective theory which demonstrates process of annihilation, we will assume that other rates than the single exciton band rates are not seriously influenced by delocalization of exciton states. We will only reflect the transformation between the local state representation and the eigenstates of the system. A convenient tool to correctly perform basis transformation, and to account for dephasing accompanying energy transfer is to write energy transfer in the so-called Lindblad form. For each rate we define a corresponding projection operator, projecting from the initial to the final state. Thus for the rate $k^{(n)}_{ef}$, which connects states $|n\rangle_2$ to states $|n\rangle_1$, we define projection operator $I_n=|n\rangle_1\langle n|_2$, and the corresponding contribution to the Lindblad form is ${\cal L}^{(I)}_n\rho = k^{(n)}_{ef} I_n\rho I_n^{\dagger} - k^{(n)}_{ef}\frac{1}{2}(I_n^{\dagger} I_n \rho + \rho I_n^{\dagger}I_n)$. Similarly, we define Lindblad form ${\cal L}^{(A)}$ for exciton-exciton annihilation, and Lindblad form ${\cal L}^{(R)}$ for the energy relaxation in the single exciton band. If single exciton states are delocalized, we also have an inverse transformation $|n\rangle_1 = \sum_{\alpha}c_{n \alpha}|\alpha\rangle$. The internal conversion rate between (localized) $f$-type state $|n\rangle_2$ and a delocalized state $|\alpha\rangle$ then reads as $k^{(n)}_{\alpha f} = \sum_{i}k^{(n)}_{ef} |c_{i\alpha}|^2$ \cite{Bruggemann2003}. Delocalization is also possible between two-exciton states, but it will not be discussed here, because we only have one two-exciton state in our dimer. 

The excited state decay can happen either radiatively, with $k_{fl}$,
or non-radiatively, with $k_{nr}$. The quantum yield of fluorescence
is given by $\phi=\frac{k_{fl}}{k_{fl}+k_{nr}}$. In our calculations
we use parameters in a range typical for (bacterio)chlorophylls,
$k_{fl}+k_{nr}=\frac{1}{3\text{ ns}}$, $\phi=0.3$. This gives $k_{fl}=\frac{1}{10\text{ ns}}$
and $k_{nr}=\frac{1}{4.3\text{ ns}}$. We consider a transition
dipole strength of our pigments to be $\left|\mu_{chl}\right|=4\text{ D}$.
The emission rate of a transition with transition dipole $\left|\mu_{i}\right|$
is scaled by its oscillator strength as $k_{fl}^{i}=\frac{\left|\mu_{i}\right|^{2}}{\left|\mu_{chl}\right|^{2}}k_{fl}$.
Because the e-e annihilation and fluorescence emission occurs on much
slower timescale than energy transfer, we can consider the one-exciton
(and also two-exciton) populations to be at thermal quasi-equilibrium
at all times during FL emission: $P_{i}(t)=\frac{e^{-\frac{e_{i}}{k_{B}T}}}{Z}P_{e}(t)$,
where $Z$ is the usual statistical sum and $P_{e}(t)$ is the overall
population of the one-exciton manifold, $P_{e}(t)=\sum_{i}P_{i}(t)$.
The decay of the one-exciton manifold is

\begin{align}
\frac{dP_{e}(t)}{dt} & =\frac{d\sum_{i}P_{i}(t)}{dt}=\sum_{i}\frac{dP_{i}(t)}{dt}=-k_{nr}\sum_{i}\frac{e^{-\frac{e_{i}}{k_{B}T}}}{Z}P_{e}(t)-\sum_{i}k_{fl}^{i}\frac{e^{-\frac{e_{i}}{k_{B}T}}}{Z}P_{e}(t)\nonumber \\
\frac{dP_{e}(t)}{dt} & =-k_{nr}P_{e}(t)-\left\langle k_{fl}^{e}\right\rangle P_{e}(t).
\end{align}

\noindent{}The one-exciton manifold thus decays with a thermally-averaged fluorescence
decay rate 

\begin{equation}
\left\langle k_{fl}^{e}\right\rangle =\sum_{i}k_{fl}^{i}\frac{e^{-\frac{e_{i}}{k_{B}T}}}{Z}
\end{equation}

\noindent{}(and the usual non-radiative rate). The fluorescence intensity is
proportional to the same rate, as is readily shown:

\begin{equation}
I_{fl}(t)=\sum_{i}k_{fl}^{i}P_{i}(t)=\sum_{i}k_{fl}^{i}\frac{e^{-\frac{e_{i}}{k_{B}T}}}{Z}P_{e}(t)=\left\langle k_{fl}^{e}\right\rangle P_{e}(t).
\end{equation}

\noindent{}Considering the two-exciton manifold, a similar effective description can be used. In our case, however, there is only one (localized) state in the doubly-excited state manifold, decaying with the sum fluorescence rate

\begin{equation}
\left\langle k_{fl}^{ee}\right\rangle =k_{fl}^{1}+k_{fl}^{2}=\frac{\left|\mu_{1}\right|^{2}+\left|\mu_{2}\right|^{2}}{\left|\mu_{chl}\right|^{2}}k_{fl}
\end{equation}

\noindent{}(and with the sum non-radiative rate $2k_{nr}$). The effective states
obey kinetic equations

\begin{equation}
\label{eq:KineticEquations}
\frac{d}{dt}\begin{pmatrix}P_{e}\\
P_{ee}\\
P_{f}
\end{pmatrix}=\begin{pmatrix}-\left\langle k_{fl}^{e}\right\rangle -k_{nr} & \left\langle k_{fl}^{ee}\right\rangle +2k_{nr} & k_{ic}\\
0 & -\left\langle k_{fl}^{ee}\right\rangle -2k_{nr}-k_{fee} & k_{eef}\\
0 & k_{fee} & -k_{eef}-k_{ic}
\end{pmatrix}\begin{pmatrix}P_{e}\\
P_{ee}\\
P_{f}
\end{pmatrix},
\end{equation}

\noindent{}and the time-resolved fluorescence intensity is

\begin{equation}
I_{fl}(t)=\left\langle k_{fl}^{e}\right\rangle P_{e}(t)+\left\langle k_{fl}^{ee}\right\rangle P_{ee}(t).
\end{equation}

\subsection*{(In)dependent molecules and pathway cancellation}


In the main text, we have discussed the presence of a cross-peak in the F-2DES spectrum. In the discussion around Fig. \ref{fig:pathways}, we have demonstrated that the magnitude of a cross peak between two states is given by pathway cancellation. We mention that the individual pathways cannot be dismissed on physical basis, as they represent merely terms in a perturbative expansion. Here we return to this point considering independent molecules.
For independent molecules, obviously, no cross-peak in the correlation 2D spectrum should occur. This requires that some cancellation between the pathways occurs, and an inspection of the Liouville pathways confirms that this indeed is indeed the case. Cancellation occurs not only for integrated fluorescence detection in which frequency of the emission is not resolved, but it appears in general on the level of precise cancellation under detection with any resolution. The SE pathway of Fig. \ref{fig:pathways} yields a photon emitted from state $|1\rangle_1$. The ESA pathway with positive sign yields two-photons, one from the state $|2\rangle_1$ and one from the state $|1\rangle_1$. The latter corresponds exactly to the SE pathway, and the former corresponds exactly to the GSB contribution from Fig. \ref{fig:pathways}. In Ref. \cite{Karki2018}, the GSB pathway is claimed to contribute with zero, because it contains interactions with two molecules which are independent of each other and cannot therefore produce a collective signal with the given phase signature. However, such claim over-interprets the response functions and asserts that the same physical criteria that apply to observables (e.g. non-existence of correlations for independent molecules) should apply to members of formal perturbative expansion of the total signal. 
The method of Feynman diagrams allows us to do a bookkeeping on various perturbative contributions. There is nothing which can justify a selective neglect of some perturbative contributions based on the non-interaction of the molecules, especially when it becomes clear that the cancellation on the level of the total signal occurs automatically.  

In this particular case, it is clear that for each of the two non-interacting molecules one can write a perturbative expansion of their wavefunction in terms of the external fields. Such expansion contains all orders of external field. The total signal is obtained by calculating the observable (e.g. population of excited states or non-linear polarization) using these perturbed wavefunctions. Individual response function, represented by various double-sided Feynman diagrams, correspond to different combinations of non-linear responses originating on the two molecules. Let us denote the total Hamiltonian operators of two molecules $1$ and $2$, including their environment, by $H_1$ and $H_2$. The two molecules do not interact, but they are positioned within the same macromolecules (e.g. photosynthetic antenna) which is much smaller in dimension than the wavelength of the light. Correspondingly, we can write their transition dipole moments as $\mathbf{\mu}_n(\mathbf{r})=\mathbf{\mu}_n\delta(\mathbf{r}-\mathbf{R})$, $n=1,2$, where $\mathbf{R}$ is the position of the macromolecule. Each of the molecules can be assigned its own wavefunction ($\psi_1(t)\rangle$ and $|\psi_2(t)\rangle$), which evolve under the unitary evolution operators $U_1(t,t^\prime)$ and $U_1(t,t^\prime)$, respectively. Applying ultrashort pulses $E_k$, $k=1,\dots,4$ with phases $\phi_1,\dots,\phi_4$ and time delays $t_1$, $t_2$ and $t_3$ already mentioned earlier, we can write 
\begin{eqnarray}
|\psi_1(t)\rangle = |\psi^{(0)}_1(t)\rangle +\sum_k|\psi_1^{(E_k)}(t)\rangle +\sum_{kl}|\psi_1^{(E_k,E_l)}(t)\rangle+\dots \nonumber \\
|\psi_2(t)\rangle = |\psi^{(0)}_2(t)\rangle +\sum_k|\psi_2^{(E_k)}(t)\rangle +\sum_{kl}|\psi_2^{(E_k,E_l)}(t)\rangle+\dots
\end{eqnarray}
where different contribution correspond different orders of perturbation in external field, for instance, 
\begin{eqnarray}
|\psi_1^{(E_1)}(t)\rangle =  U_{1}(t,0)|e_1\rangle\mathbf{\mu}_{eg}\langle g_1|g_1\rangle |\xi_{\rm{eq}}\rangle \nonumber \\
|\psi_1^{(E_4)}(t)\rangle =  U_{1}(t, t_1+t_2+t_3)|e_1\rangle\mathbf{\mu}_{eg}\langle g_1|U_{1}(t_1+t_2+t_3)|g_1\rangle |\xi_{\rm{eq}}\rangle.
\end{eqnarray}
The total wavefunction of the system of two independent molecules and their environments reads as
\begin{eqnarray}
|\psi(t)\rangle = |\psi_1(t)\rangle|\psi_2(t)\rangle & = & \left(|\psi^{(0)}_1(t)\rangle +\sum_k|\psi_1^{(E_k)}(t)\rangle + \dots \right)\left( |\psi^{(0)}_2(t)\rangle +\sum_l|\psi_2^{(E_l)}(t)\rangle + \dots \right) \nonumber \\
& = & \dots + \sum_{kl} |\psi_1^{(E_k)}(t)\rangle|\psi^{(E_l)}_2(t)\rangle + \dots,
\end{eqnarray}
where on the second line, we singled out a contribution resulting from a product of singly perturbed wavefunctions. When we calculate populations of excited states after interaction with four pulses, we get (among others) the following probability of joint population of excited states of both molecules
\begin{eqnarray}
\label{eq:P12}
P_{12}  &=& \langle\psi(t)|e_1\rangle |e_2\rangle\langle e_2|\langle e_1|\psi(t)\rangle \nonumber \\
&=& \dots + \langle \psi^{(E_2)}|e_1\rangle \langle \psi^{(E_3)}|e_2\rangle \langle e_1|\psi^{(E_1)}\rangle \langle e_2|\psi^{(E_4)}\rangle + \dots.    
\end{eqnarray}
On the second line of Eq. (\ref{eq:P12}) we singled out a contribution corresponding to the ESA2 pathway. For independent molecules $1$ and $2$, existence of such joint probability is perfectly normal, as is the existence of the joint probability of molecule $2$ being excited, while molecule $1$ is in the ground state which reads 
\begin{eqnarray}
\label{eq:P02}
P_{02}  &=& \langle\psi(t)|g_1\rangle |e_2\rangle\langle e_2|\langle g_1|\psi(t)\rangle \nonumber \\
&=& \dots + \langle \psi^{(0)}_1|g_1\rangle \langle \psi^{(E_3)}_2|e_2\rangle \langle g_1|\psi^{(E_1,E_2)}_1\rangle \langle e_2|\psi^{(E_4)}_2\rangle + \dots.    
\end{eqnarray}
Here, we singled out the contribution corresponding to the GSB diagram in Fig. \ref{fig:pathways}. Again there is no reason for an assumption that this is not a valid contribution to the joint probability. It is a product of non-zero terms, and there is no need for any dependence of the two molecules $1$ and $2$ for such products to be non-zero. 

\subsection*{The numerical simulation: procedure and parameters}

In this section we describe the parameters and procedure for the numerical simulations presented in the main text. The model effective dimer has the following parameters summarized in Tab. \ref{tab:SysParams} (unless explicitly mentioned otherwise). Transition energies: $\hbar\omega^{(1)}_{eg}=12000$ cm$^{-1}$, $\hbar\omega^{(2)}_{eg}=12500$ cm$^{-1}$, $\hbar\omega^{(1)}_{fe}=\hbar\omega^{(2)}_{fe}=13500$ cm$^{-1}$. With these energy gaps there is virtually no delocalization between the two-exciton and higher excited states for all values of resonance coupling considered. Effective transfer rate from two-exciton $|ee\rangle$ states to the higher excited  $|f\rangle$ states  $k_{fee}=\frac{1}{10\text{ ps}}$, intramolecular relaxation rate between $|n\rangle_2 \rightarrow |n\rangle_1$  $k_{ic}=\frac{1}{100\text{ fs}}$, and the nonradiative relaxation rate $k_{r}=\frac{1}{4.3\text{ ns}}$ and the radiative relaxation rate $k_{fl}=\frac{1}{10\text{ ns}}$. The radiative rate scales with the transition oscillator strength and for the one- and double-exction manifolds an effective rate is calculated, as described in the section on system dynamics above. In our effective dimer the two (groups of) molecules have transition dipoles $\mu^{(1)}_{eg}=\frac{4}{\sqrt{10}}\left(3,1,0\right)$, $\mu^{(2)}_{eg}=\frac{3.3}{\sqrt{10}}\left(3,-1,0\right)$ in the $xy$ plane and their distance in the z plane is varied from 50 \AA\: to 6 \AA\: in order to change the coupling, which is calculated in the dipole-dipole approximation. The transition dipoles to the higher excited states are taken to be proportional to the ground-one exciton transitions, $\mu^{(n)}_{fe}=\nu\mu^{(n)}_{eg}$. As typical values of $\nu$ are on the order of 1, we take $\nu=1$ for the calculation. The coupling between the $g\rightarrow e$ and $e \rightarrow f$ are calculated in the same way as between the single-exciton states, $J_{e_{n}g,f_{m}e_{m}}=\nu J_{e_{n}g,e_{m}g}$, highlighting the fact that it is the \textit{transitions} which are coupled, rather than the states. The e-e annihilation process consists of exciton migration producing spatially co-localized two excitation, population transfer from two-exciton $|ee\rangle$ state to higher excited $|f\rangle$ state, with following internal conversion to the $|e\rangle$ state. In our effective description we include the exciton migration time into the $k_{fee}$ rate. The backward $k_{eef}$ rate is neglected, as long as it is slower than $k_{ic}$ it makes no difference. The molecule environment is described by an overdamped Brownian oscillator spectral density\cite{Mukamel1995Book}, with correlation time 100 fs, reorganization energy $\lambda=50$ cm$^{-1}$, at temperature of 300K. The $|f\rangle$ states have an additional dephasing given by the 100 fs relaxation rate. The calculations were done using the Quantarhei package\cite{quantarhei2018} which computes lineshapes by cumulant expansion. The 2D spectra at time zero are calculated by response function formalism, using the Aceto library\cite{aceto2018}, with all-parallel polarization and orientational averaging (see the enclosed scripts). The standard 2DES is calculated as usual, by summing the individual Liouville space pathways $S_{2DES}=GSB+SE-ESA$. For the calculation of the F-2DES, the system state after the interactions with the four pulses presents an initial condition for the dynamics during which the fluorescence signal is emitted. First, the kinetic equations Eq. \eqref{eq:KineticEquations} are solved numerically, keeping track of the the initial condition (solving for $P_e=1,\;P_{ee}=1,\;P_f=1$ separately). The fluorescence emission signal is calculated as $I_{FL}(t)=\left\langle k_{FL}^{e}\right\rangle P_{e}(t)+\left\langle k_{FL}^{ee}\right\rangle P_{ee}(t)$. Then the individual pathways are summed, weighted by their contribution to the signal obtained for the initial condition which corresponds to the state which the pathways ends in. By this procedure, the calculation cost for the F-2DES is only increased by the time it takes to integrate Eq. \eqref{eq:KineticEquations}. 

\noindent{}We would like to emphasize that the calculation described here and in the section on the system dynamics above is rather general, not bound to the effective dimer model we use for the calculations. It can be applied to any multichromophoric system, as long as the excitation equilibration within the one- and double- exciton manifolds is faster than the radiative and nonradiative interband relaxation.       

\subsection*{The cross-peak amplitude}
\subsubsection*{Collective vs local basis}

In this section we derive the dependence of the cross-peak on the coupling and annihilation. We consider a simplified version of the dimer used for the calculations in the main text, consisting of two coupled two-level molecules with states $\left|g\right\rangle ,\left|e\right\rangle $. 
The local basis description for the system is

\begin{align}
\mathbb{1} & =\mathbb{1}_{1}\otimes\mathbb{1}_{2}\nonumber \\
\mathbb{1} & =\left(\left|g_{1}\right\rangle \left\langle g_{1}\right|+\left|e_{1}\right\rangle \left\langle e_{1}\right|\right)\otimes\left(\left|g_{2}\right\rangle \left\langle g_{2}\right|+\left|e_{2}\right\rangle \left\langle e_{2}\right|\right)
\end{align}

\noindent{}We can always formally introduce collective states $\left|0\right\rangle =\left|g_{1}\right\rangle \left|g_{2}\right\rangle $,
$\left|1\right\rangle =\left|e_{1}\right\rangle \left|g_{2}\right\rangle $,
$\left|2\right\rangle =\left|g_{1}\right\rangle \left|e_{2}\right\rangle $,
$\left|12\right\rangle =\left|e_{1}\right\rangle \left|e_{2}\right\rangle $.
These also form a basis,

\begin{equation}
\mathbb{1}=\left|0\right\rangle \left\langle 0\right|+\left|1\right\rangle \left\langle 1\right|+\left|2\right\rangle \left\langle 2\right|+\left|12\right\rangle \left\langle 12\right|.
\end{equation}

\noindent{}The system Hamiltonian is (shifting the ground state energy to zero): 

\begin{align}
H & =H_{1}+H_{2}+H_{12},\nonumber \\
H_{1}+H_{2} & =\left|0\right\rangle 0 \left\langle 0\right|+\left|1\right\rangle e_1\left\langle 1\right| +\left|2\right\rangle e_2\left\langle 2\right|+\left|12\right\rangle \left(e_1+e_2\right)\left\langle 12\right|,\nonumber \\
H_{12} & =\left|1\right\rangle J_{12}\left\langle 2\right|+\left|2\right\rangle J_{12}\left\langle 1\right|.
\end{align}

\noindent{}To have distinguishable molecules, we will consider $e_{1}\ne e_{2}$.
When the coupling $J$ is weak and treated perturbatively, we can
use either the local or collective basis, these descriptions being
equivalent. However, when the coupling $J$ is strong, the collective
eigenstates of the system have to be used.

\subsubsection*{2DES}

Let us calculate a traditional 2DES cross-peak for interacting
molecules. To restrict the number of pathways needed, without loss
of generality, let us focus on the rephasing part of the $\left(\omega_{1},\omega_{3}\right)=\frac{1}{\hbar}\left(e_1, e_2 \right)$
cross peak at the waiting time zero. The Liouville-space pathways that contribute
are in Fig. \ref{fig:pathways} in the main text.
We wrote the pathways in the collective states, as this description is general. 
For weakly-interacting pigments with localized excitation, also the local basis can be used. In this view,
in the first pathway both molecules get bleached and in the second
pathway one first molecule gets excited and the second one gets bleached.
Because the coupling is weak, the molecule states and thus pathways
are independent; there is therefore no part of the system with 'imprinted'
phase from all three pulses. As a result, there is no cross peak,
as neither of the pathways contributes. In the collective states, both
pathways contribute, but with a different sign (1st pathway has $+1$,
second $-1$, as seen from number of interactions from the right).
The pre-factors are the same, $|\mu_{e_{1}}|^{2}|\mu_{e_{2}}|^{2}$.
So the pathways cancel out and no cross peak appears.\footnote{We did not consider pathways with a coherence in population time,
as for independent molecules this coherence dephases rapidly. The
same cancellation, however, applies also to these, with the GSB-type
pathway replaced by a SE-type pathway. }

The utility of the collective state description becomes apparent when
slowly increasing the coupling $J$ (for instance by bringing the
molecules closer to each other). The system eigenstates interacting
with light become linear combinations of the molecular states,

\begin{align}
\left|1\right\rangle  & =sin\vartheta\left|e_{1}\right\rangle \left|g_{2}\right\rangle +cos\vartheta\left|g_{1}\right\rangle \left|e_{2}\right\rangle \nonumber \\
\left|2\right\rangle  & =cos\vartheta\left|e_{1}\right\rangle \left|g_{2}\right\rangle -sin\vartheta\left|g_{1}\right\rangle \left|e_{2}\right\rangle \nonumber \\
\left|0\right\rangle  & =\left|g_{1}\right\rangle \left|g_{2}\right\rangle ,\left|12\right\rangle =\left|e_{1}\right\rangle \left|e_{2}\right\rangle .
\end{align}

\noindent{}The mixing angle is given by the coupling and energy gap, $tan2\vartheta=\frac{2J}{e_{2}-e_{1}}$.
The transition dipoles are also transformed, 

\begin{align}
\mu_{2,0} & =sin\vartheta\mu_{e_{1}}+cos\vartheta\mu_{e_{2}}\nonumber \\
\mu_{1,0} & =cos\vartheta\mu_{e_{1}}-sin\vartheta\mu_{e_{2}}\\
\mu_{21,2} & =sin\vartheta\mu_{e_{2}}+cos\vartheta\mu_{e_{1}}\nonumber \\
\mu_{21,1} & =cos\vartheta\mu_{e_{2}}-sin\vartheta\mu_{e_{1}}
\end{align}

\noindent{}Apparently, using the local basis is no longer possible, as the light
excites superpositions of the molecular states. However, the collective
states can be used the same as before.

The only difference are the pre-factors for the pathways: the GSB pathway
has a pre-factor $|\mu_{1,0}|^{2}|\mu_{2,0}|^{2}$, while the ESA pathway
has a pre-factor $|\mu_{1,0}|^{2}|\mu_{21,1}|^{2}$. The cross-peak
cancellation is then not complete: taking for simplicity $\mu_{e_{1}}=\mu_{e_{2}}=\mu$,
we get

\begin{align}
GSB-ESA & \propto|\mu_{1,0}|^{2}\left\{ |\mu_{2,0}|^{2}-|\mu_{21,1}|^{2}\right\} \label{eq:trad2DES}\\
GSB-ESA & \propto|\mu|^{4}\left(-sin\vartheta+cos\vartheta\right)^{2}\times\nonumber \\
 & \left\{ \left(sin\vartheta+cos\vartheta\right)^{2}-\left(-sin\vartheta+cos\vartheta\right)^{2}\right\} \nonumber \\
 & =|\mu|^{4}\left(1-sin2\vartheta\right)2sin2\vartheta.
\end{align}

\noindent{}For small $J$ (compared to the energy gap) we have $\left(1-sin2\vartheta\right)\approx1$
and $sin2\vartheta\approx tan2\vartheta=\frac{2J}{e_{2}-e_{1}}.$
The magnitude of the cross peak is therefore directly proportional
to the coupling between the molecules and the cross peak is a good signature of delocalization. Formulation in the collective states enables
us to increase smoothly the coupling from $J=0$ and observe the appearance
of the cross peak.

\subsubsection*{F-2DES}

There are three main differences between the F-2DES and the traditional 2DES. 
First, there
is one more interaction with the light field, which means one more
commutator with $\mu$, giving rise to one more contributing Liouville
space pathway. Second, for the FL to appear, the system has to end up in the excited state
after the interaction with the pulses. And third, the evolution of
the system after the interaction with the four pulses matters, as
during this evolution the FL signal is emitted. The
pathways contributing to the cross peak are presented in Fig. \ref{fig:pathways}.
The first two pathways are analogical to the ones in the
traditional 2DES, but now they have the same sign, so they do not
cancel. The last ESA-type pathway has an opposite sign, and it is special
in that it ends in the state with two excitations in the system. The
cross-peak amplitude will be proportional to:

\begin{equation}
-GSB-ESA1+ESA2=-|\mu_{1,0}|^{2}|\mu_{2,0}|^{2}+\left(\Gamma-1\right)|\mu_{1,0}|^{2}|\mu_{21,1}|^{2},
\end{equation}

\noindent{}where by $\Gamma$ we denote the contribution of the doubly-excited
pathway, considering time-integrated detection. $\Gamma$ is given by the emission rate of the double excited states, $k^{ee}_{fl}$, relative to $k^{e}_{fl}$, and by the exciton-exciton annihilation rate. In this section we will consider the case of weakly-coupled pigments (no significant delocalization and redistribution of oscillator strength) and of equal oscillator strength. In the case that both excitations survive and independently contribute
to the signal (no annihilation), we have $\Gamma=2$. In the case that there
is an efficient annihilation (as is typically the case of molecules),
we have $\Gamma=1$. Actually, already from the expression above we
can see what is the effect of $\Gamma$. For no annihilation, $\Gamma=2$,
we get exactly the same as in the traditional 2DES, Eq. (\ref{eq:trad2DES}),
only with opposite overall sign. Then there is no cross-peak present. In contrast, for full annihilation, $\Gamma=1$, we get
only the GSB pathway contributing. As a result, there is a cross peak
even for practically independent molecules, as long as the presence
of two excitations results in annihilation. Concerning the sign of
the cross peak, it is the same as for the diagonal peak $\left(\omega_{1},\omega_{3}\right)=\left(e_{1},e_{1}\right)$.

As before, we can write explicitly the magnitude of the cross peak (again
for $\mu_{e_{1}}=\mu_{e_{2}}=\mu$):

\begin{equation}
-GSB-ESA1-ESA2\propto|\mu|^{4}\left(1-sin2\vartheta\right)\left\{ \Gamma-2-\Gamma sin2\vartheta\right\} .
\end{equation}

\noindent{}For $\Gamma=2$ the peak magnitude is, again, proportional to the
coupling. However, for any $\Gamma<2$, there is a non-zero cross-peak
amplitude even in the limit of $J\rightarrow0$. In molecules, there
is a significant range of coupling magnitude for which there is no
delocalization, but efficient annihilation (essentially the F\"orster
regime). If we denote the extent of annihilation $A=2-\Gamma$ ($A=1$
for full annihilation and $A=0$ for no annihilation), and consider
still the coupling much smaller than the energy gap, $sin2\vartheta\approx0$,
the FL-detected 2DES cross peak is directly proportional to the annihilation,

\begin{equation}
-GSB-ESA1-ESA2\propto-|\mu|^{4}A.
\end{equation}

\noindent{}For larger couplings the unequal radiative power, delocalization and annihilation all contribute to the cross-peak.

\bibliography{Mendeley,fd2d-pm}

\end{document}